\documentclass[a4paper]{report}
\usepackage[utf8]{inputenc}
\usepackage[T1]{fontenc}
\usepackage{RJournal}
\usepackage{amsmath,amssymb,array,mathtools}
\usepackage{booktabs}
\usepackage{multicol}
\usepackage{placeins}
\usepackage{float}
\usepackage[ruled,vlined]{algorithm2e}

\begin{document}

\sectionhead{}
\volume{XX}
\volnumber{YY}
\year{2020}
\month{October}

\begin{article}
\title{False Discovery Rate Computation: \\Illustrations and Modifications}
\author{by Megan Hollister Murray and Jeffrey D. Blume}

\maketitle

\abstract{
False discovery rates (FDR) are an essential component of statistical inference, representing the propensity for an observed result to be mistaken. FDR estimates should accompany observed results to help the user contextualize the relevance and potential impact of findings. This paper introduces a new user-friendly R package for computing FDRs and adjusting p-values for FDR control \citep{R}. These tools respect the critical difference between the adjusted p-value and the estimated FDR for a particular finding, which are sometimes numerically identical but are often confused in practice. Newly augmented methods for estimating the null proportion of findings - an important part of the FDR estimation procedure - are proposed and evaluated. The package is broad, encompassing a variety of methods for FDR estimation and FDR control, and includes plotting functions for easy display of results. Through extensive illustrations, we strongly encourage wider reporting of false discovery rates for observed findings. 
}

\section{Introduction}

The reporting of observed results is not without controversy when multiple comparisons or multiple testing is involved. Classically, p-values were adjusted to maintain control of the family-wise error rate (FWER). However, this control can come at the cost of substantial Type II Error rate inflation, especially in large-scale inference settings where the number of comparisons is several orders of magnitude greater than the sample size. Large scale inference settings occur frequently in the analysis of genomic, imaging, and proteomic data, for example. Recently, it has become popular to control the false discovery rate (FDR) instead of the FWER in these settings because its Type II Error rate inflation is much less severe. The FDR is essentially the propensity for a finding to be mistaken i.e., the propensity for a non-null claim to be, in fact, wrong. 

Controlling the FDR at or below a specific level, say $\gamma$, does \textit{not} imply that the Type I Error rate, per-comparison or family-wise, is also controlled at the same level. The increase in Type I Errors that is allowed by FDR control is accompanied by fewer Type II errors. Moreover, different approaches to controlling the FDR allow for different degrees of error tradeoff. And software for implementing these approaches vary widely in their scope, options, and accessibility. In addition, methods for controlling the FDR, which use the classical rejection-testing framework, are often confused with the methods used to provide an estimate of the FDR for a particular result.

The \texttt{FDRestimation} package distinguishes between methods for FDR control and methods for FDR estimation, and it allows the user to easily access complex statistical routines for computing the desired quantity. The plotting functions allow users to visually assess results and differences between methods. We should note that the base package function \texttt{stats::p.adjust}  is now frequently used to compute the estimated FDR, however \texttt{stats::p.adjust} actually reports the adjusted p-values for FDR control, and these are not always the same thing. More on this important distinction later. Our package also provides a wide range of methods for estimating the FDR, estimating the proportion of null results, and computing the adjusted p-values. We hope by clearly illustrating the usage of our package in routine settings that these FDR methods will become more accessible and gain even more popularity in routine practice. 

\subsection{Simple Motivating Example}

We begin with a simple example to fix ideas. Table \ref{tab1} shows five unadjusted (raw) p-values for experimental features along with their corresponding Z-values. The third column lists the Benjamini-Hochberg adjusted p-values to be used for FDR control \citep{bh:1995}. Controlling the FDR at level $\gamma$ amounts to selecting all of the adjusted p-values in column 3 that are below $\gamma$. Note here that the adjusted p-values are monotonically increasing, just like the raw p-values, but inflated. 

If the goal is to control the FDR at 5$\%$, then only the first feature would be declared interesting and selected. Throughout the paper, we use the term ``interesting" to describe features that are selected by a procedure with FDR control. We do not use the term ``significant" in order to avoid confusion with those features that would have been selected from by a procedure with strict Type I Error control.

\begin{table}[!t]
\centering
\begin{tabular}{|l|l|l|l|l|l|}
\hline
\multicolumn{1}{|l|}{\bf{Feature}}    & \multicolumn{1}{l|}{\bf{Raw p-value}}  & \multicolumn{1}{l|}{\bf{Z-value}} & \multicolumn{1}{l|}{\bf{Adjusted p-value}} & \multicolumn{1}{l|}{\bf{FDR}} & \multicolumn{1}{l|}{\bf{Lower Bound FDR}}\\ \hline

\multicolumn{1}{|l|}{Feature 1}     &\multicolumn{1}{l|}{0.005}     & \multicolumn{1}{l|}{2.807} & \multicolumn{1}{l|}{0.025} & \multicolumn{1}{l|}{0.025} & \multicolumn{1}{l|}{0.019}\\ \hline

\multicolumn{1}{|l|}{Feature 2}     &\multicolumn{1}{l|}{0.049}     & \multicolumn{1}{l|}{1.969} & \multicolumn{1}{l|}{0.064} & \multicolumn{1}{l|}{0.122} & \multicolumn{1}{l|}{0.126}\\ \hline

\multicolumn{1}{|l|}{Feature 3}     &\multicolumn{1}{l|}{0.050}     & \multicolumn{1}{l|}{1.960} & \multicolumn{1}{l|}{0.064} & \multicolumn{1}{l|}{0.083} & \multicolumn{1}{l|}{0.128}\\ \hline

\multicolumn{1}{|l|}{Feature 4}     &\multicolumn{1}{l|}{0.051}     & \multicolumn{1}{l|}{1.951} & \multicolumn{1}{l|}{0.064} & \multicolumn{1}{l|}{0.064} & \multicolumn{1}{l|}{0.130}\\ \hline

\multicolumn{1}{|l|}{Feature 5}     &\multicolumn{1}{l|}{0.700}     & \multicolumn{1}{l|}{0.385} & \multicolumn{1}{l|}{0.700} & \multicolumn{1}{l|}{0.700} & \multicolumn{1}{l|}{0.481}\\ \hline                                                                                                                                                                                                                                                                             
\end{tabular}
\caption {\label{tab1} Example with 5 features using the Benjamini-Hochberg adjustment and assuming a two-sided normal distribution. }
\end{table}
\FloatBarrier

The fourth column presents FDR estimates for each feature. As we show later, there are several ways to invert the FDR control procedures to yield an estimate of the FDR. Our package performs this inversion for most popular methods. The FDRs here were obtained by inverting the Benjamini-Hochberg FDR control procedure, and so we will refer to them as the BH FDRs \citep{bh:1995}. In practice we find these estimates to be the most context useful when making scientific decisions about which findings to pursue.

Importantly, these are clearly not identical to the BH adjusted p-values nor are they even monotone. The non-monotonicity results from the group-wise p-value adjustment procedure ("step-up") and the non-smooth estimate of the p-value mixture distribution, which is needed for FDR estimation. The important insight is that the set of features that are selected by the FDR control procedure is not equivalent to the set of feature whose individual FDR is less than the control threshold. For example, if the FDR threshold was $\gamma$=0.07, then the first 4 features would be selected by BH to control the group-wise FDR at 7$\%$. However, only the first and fourth features have estimated false discovery rates below 0.07, and thus only these two features would be reported as having a false discovery propensity less than 7$\%$. Note that both approaches come from the same Benjamini-Hochberg machinery, and thus have the same distributional assumptions. The distinction between adjusted p-values and estimated FDRs are critical here. 

Because FDRs are only estimates, and because there are a variety of estimation approaches, it helps to have a feature-specific benchmark for each FDR. The fifth column provides such a benchmark; it displays a well-known lower bound on the FDR assuming a gaussian posterior and a null proportion of 50$\%$ These assumptions are relatively benign for reasons we discuss later and represent a ``best-case" scenario. This benchmark shows two things: (1) the adjusted p-values are a poor substitute for the FDRs, and (2) the smoothness of the FDR estimation approach is important.

\section{FDR Methods}

\subsection{p-value Based Approaches}

Let $p_1,...,p_m$ be the individual unadjusted p-values derived for each of m different features or tests. For clarity, the $i^{th}$ p-value is for the $i^{th}$ feature and has not been adjusted for any multiple testing. It is sometimes referred to as the ``univariate" p-value. The sorted or ranked p-values are represented by $p_{(1)},...,p_{(m)}$ where $p_{(1)}$ is the smallest, $p_{(m)}$ is the largest and with $p_{(k)}$ is the $k^{th}$ ranked p-value. 

Let $\gamma$ be the false discovery rate threshold for interesting findings. This threshold is context specific, and is either set by the researcher or according to a community standard. This threshold is specified a priori when performing FDR control procedures, but it need not be specified for FDR estimation procedures. The Benjamini-Hochberg algorithm for FDR control is to find the largest index, say $k$, such that

\begin{equation} \label{eq1}
\begin{split}
p_{(i)} \leq \gamma \frac{i}{m}  \text{ for }  i\in \{1,2,...,m\}
\end{split}
\end{equation}

This can be written compactly $k=\max{\left[i: p_{(i)} \leq \gamma i / m\right]}$. Then all features with $p_{(1)},...,p_{(k)}$ are deemed interesting at the FDR $\gamma$ threshold and considered ``findings". This is called a ``step-up" procedure because not all of the rejected features will have unadjusted p-values that meet the above criterion. Only the largest of them must meet that criterion. Because this is a ``step-up" procedure, the adjusted p-values will depend on the raw p-values from other features. The Benjamini-Hochberg adjusted p-value for the $i^{th}$ feature is notated in this paper by $\tilde{p}_{i}$ and defined in Equation \eqref{eq2}, where $\coloneqq$  means ``is defined as". 

\begin{equation} \label{eq2}
\begin{split}
\tilde{p}_{(i)} \coloneqq \min_{j \geq i}\left(\frac{p_{(j)}m}{j}\right) \leq \gamma 
\end{split}
\end{equation}

These adjusted p-values are monotone increasing in raw p-value ranking, so one can directly compare $\tilde{p}_{i}$ to $\gamma$ to see if a particular feature would be rejected as null for the FDR threshold $\gamma$. Importantly, the feature specific FDR estimates need not be monotone. To see this, re-arrange Equation \eqref{eq1} as follows in Equation \eqref{eq3}.

\begin{equation} \label{eq3}
\begin{split}
FDR_i \coloneqq \frac{p_i m}{\text{rank}(p_i)} \cdot \hat{\pi}_0
\end{split}
\end{equation}

The derivation of FDR is described in the following section. This shows that the BH procedure is, in effect, estimating the feature specific FDR as $FDR_i$. See also Efron LSI for motivation for this definition \citep{efron:2013}. Because estimation of the feature specific FDR does not include group-wise control of the FDR, the ``step-up" monotonicity condition does not apply. Thus, feature specific FDR estimates such as $FDR_i$ are not always monotone in raw p-value ranking. 

A consequence of this dichotomy is that an individual feature may be rejected at FDR $\gamma$ level by the BH algorithm even though its feature specific FDR estimate is actually greater than $\gamma$. This is largely a consequence of the smoothness of the FDR estimates and the fact that they can have substantial variability. Note that there are several methods for estimating the FDR, and some methods may be better suited to certain contexts. Our package offers several methods for FDR estimation, as described in later sections of this paper.

\begin{figure}[H]
\centering
  \includegraphics[width=0.9\linewidth]{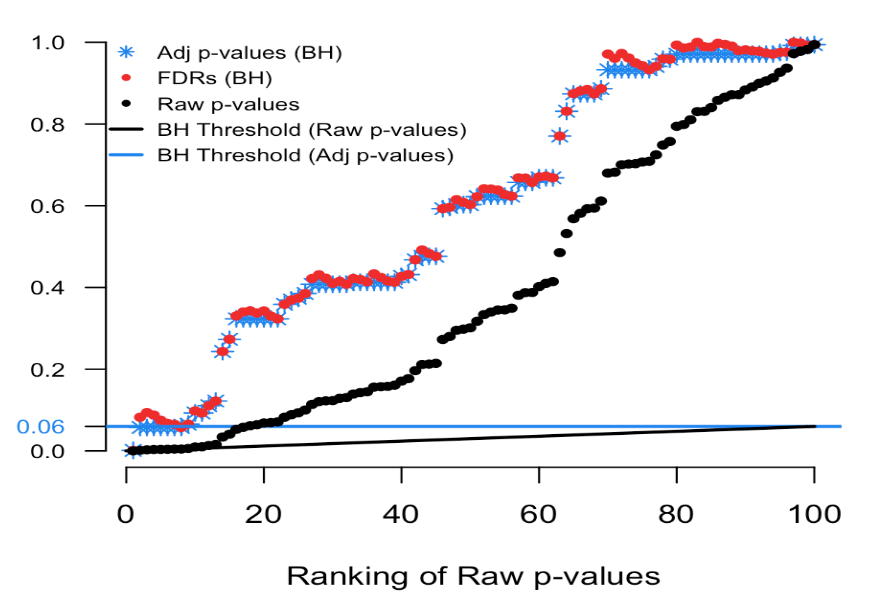}
  \caption{Simulated Example comparing raw p-values and the threshold of interest.}
  \label{fig1}
\end{figure}
\FloatBarrier

\begin{figure}[H]
\centering
  \includegraphics[width=0.9\linewidth]{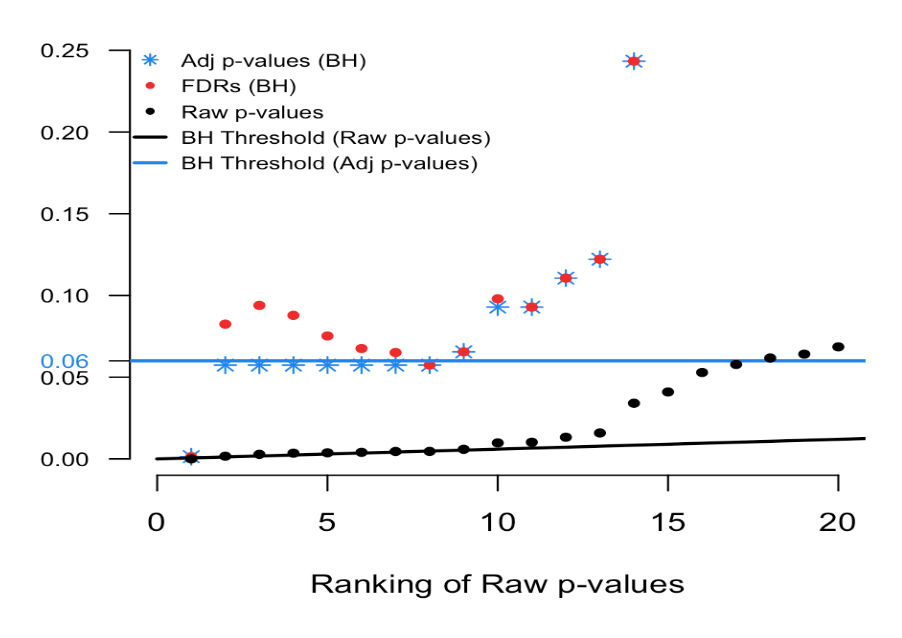}
  \caption{Magnified Graph of the simulated example.}
  \label{fig2}
\end{figure}

To illustrate we simulated real data from 100 hypothesis tests and captured the 100 raw p-values. For context, 80 of these p-values were generated from a uniform distribution (and hence under the null) while the other 20 were generated from a skewed distribution representing the alternative. Results are computed using our \texttt{p.fdr}  function, which we detail later. The raw p-values are displayed in Figure \ref{fig1} as black points; Figure \ref{fig2} shows only the 20 features with the smallest ranked raw p-values. The black sloped line is the BH rejection line from Equation \eqref{eq1}. Also included in the plot are the BH adjusted p-values (blue stars), the BH FDR threshold for interesting findings (blue horizontal line), and the BH FDR estimates (red points).

In Figure \ref{fig2} we see that exactly 8 of the adjusted p-values fall below our threshold of interest (blue line, set here to 0.06). Therefore, the BH FDR procedure that controls the group-wise FDR identifies the 8 smallest p-values as interesting findings. However, notice the non-monotonicity of the individual FDRs. Only the first and last of the 8 lowest FDRs are less than 0.06. 

\FloatBarrier

From these results it should be clear that the feature-specific FDRs and the BH adjusted p-values have different purposes and interpretations. To emphasize, when a feature is identified as 'interesting' by an FDR control procedure, it does not always follow that the feature's individual propensity to be a false discovery is less than the desired threshold. Both quantities must be computed, as the tasks are not always exchangeable.  

\subsection{Z-value Based Approaches }

For FDR estimation, it is often helpful to transform the p-values $p_{1},...,p_{m}$ to into Z-values $z_{1},...,z_{m}$ using the standard normal quantiles. For example, $z_{i}=\Phi^{-1}(1-p_{i})$ for one-sided p-values or $z_{i}=\Phi^{-1}(1-p_{i}/2)$ for two-sided p-values.  Efron explains the rationale as an attempt to leverage the distributional properties of a set of gaussian random variables\citep{efron:2013}. Note that these Z-values are not intended to be the original test statistics. We will adopt Dr. Bradley Efron's  formulation as described here \citep{efron:2013}.

We begin with the classic two-group model, which assumes each of the $m$ features is either null (distribution known) or alternative (distribution unspecified), but that this status is unknown. As a group the combined data can be used to provide an estimate of the mixture distribution, where the mixing proportion ($\pi_0$) is also unknown. Let $f_0(z)$ be the probability density function of the $z$-values when they come from the true null distribution and $f_1(z)$ be the probability density function of the $z$-values when they come from the alternative distribution. Then $F_0(\cdot)$ and $F_1(\cdot)$ denote the probability of rejection for any subset $\mathcal{Z}$ of the real line such that, 

\begin{equation} \label{eq4}
\begin{split}
F_0(\mathcal{Z})=\int_{\mathcal{Z}}f_0(z)dz \text{   and    } F_1(\mathcal{Z})=\int_{\mathcal{Z}}f_1(z)dz
\end{split}
\end{equation}

With mixing or null proportion $\pi_0$, the proportion of non-null features is simply $\pi_1=1-\pi_0$. The mixing distribution function is
 
\begin{equation} \label{eq5}
\begin{split}
F(\mathcal{Z})=\pi_0F_0(\mathcal{Z})+\pi_1F_1(\mathcal{Z})
\end{split}
\end{equation}

When working with Z-values, it is reasonable to use a gaussian distribution for the theoretical null probability density function, so that $f_0(z)\sim N(0,1)$ \citep{efron:2013}. When estimating the FDR, is it also common to assume that $\pi_0=1$ because doing so results in a conservative estimate of the FDR. Then, an application of Bayes famous theorem yields:


\begin{equation} \label{eq6}
\begin{split}
FDR(\mathcal{Z}) \coloneqq  Pr\{null | z \in \mathcal{Z}\}=\frac{\pi_0F_0(\mathcal{Z})}{F(\mathcal{Z})}
\end{split}
\end{equation}

Substituting the natural empirical estimate of the mixture distribution $F(\mathcal{Z})$ results in empirical Bayes estimates the global FDR Equation \eqref{eq6} \citep{bh:1995} \citep{efron:2013}. For example, the obvious empirical estimate of the mixing distribution function is the step function $\hat{F}(\mathcal{Z}_i)=\text{rank}(p_i)/m$. Notice that the right hand side of Equation \eqref{eq1} then looks like $\gamma \cdot \hat{F}(\mathcal{Z}_i)$ or $\gamma$ times the step function.  In some settings smoothing $\hat{F}(\mathcal{Z}_i)$ can be beneficial. Very often it is assumed $\pi_0=1$ and $F_0(\mathcal{Z})=1-\Phi(\mathcal{Z})$ for one-sided tests. An advantage of estimating the FDR from the right hand side of Equation \eqref{eq6} is that one only needs to accurately estimate the mixture distribution function to get good estimates of the FDR and this does not require the independence of the z-values. 

\begin{figure}[H]
\centering
  \includegraphics[width=0.9\linewidth]{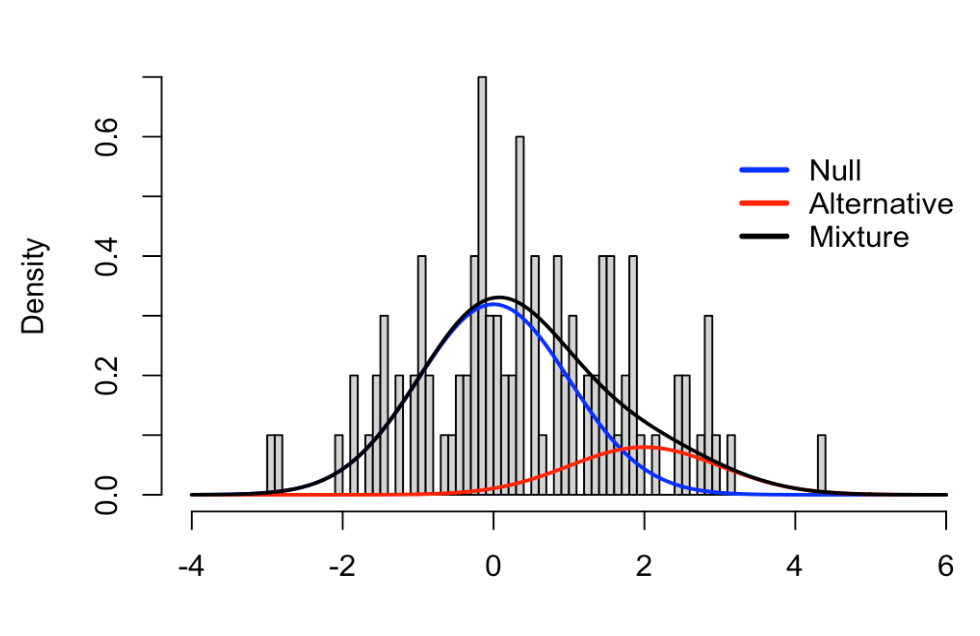}
  \caption{Simulated Example Density Histogram.}
  \label{fig3}
\end{figure}
\FloatBarrier

\begin{figure}[H]
\centering
  \includegraphics[width=0.9\linewidth]{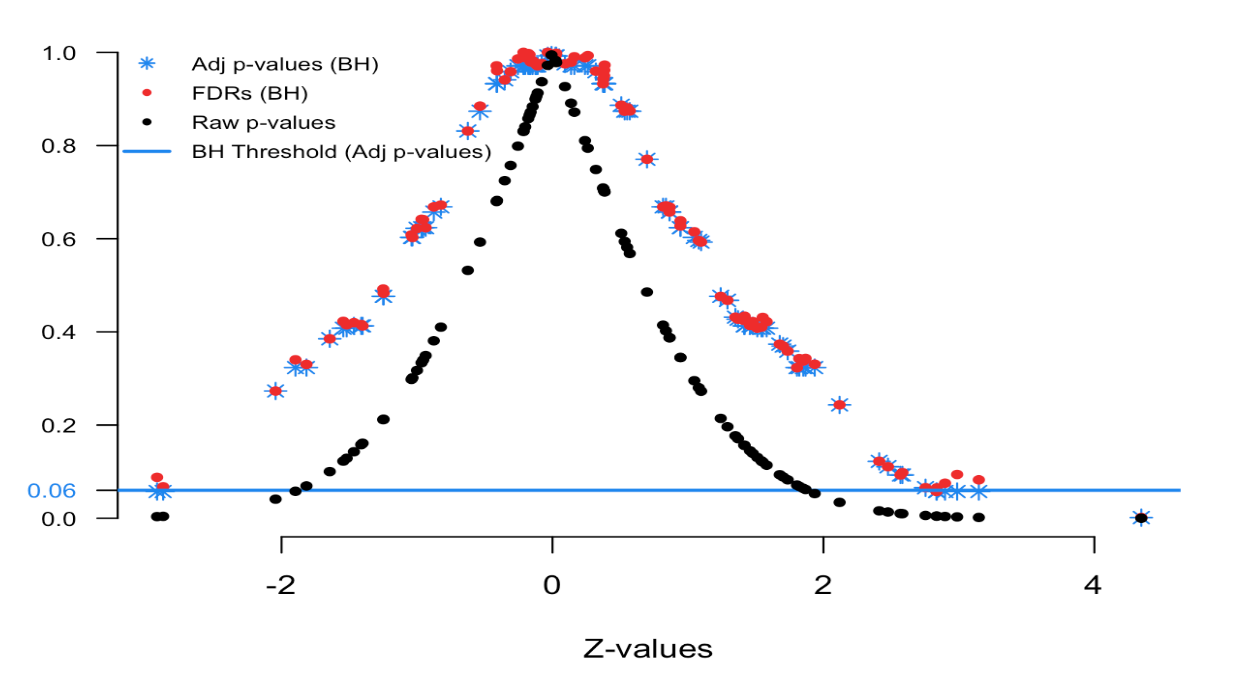}
  \caption{FDR Z-values plot}
  \label{fig4}
\end{figure}

Figures \ref{fig3} and \ref{fig4} show the application of this framework using the same simulated data as in the last example (100 tests, 80 truly null). In the z-space, the null distribution is now the standard normal and the alternative distribution was set to $N(2,1)$ (of course this is unknown, in practice). Figure \ref{fig3} shows these densities overlaid on a histogram of the raw data. The blue curve indicates the null density, the red curve indicates the alternative density, and the black curve is the mixture density with $\pi_0=0.8$. Clearly, the blue curve does not fit the histogram well, with a much shorter right tail than the histogram shows. So, assuming all 100 tests come from the null distribution does come with a penalty. 

Figure \ref{fig4} displays the relationship between the Z-values and various FDR quantities. The black dots show the raw p-values (y-axis) versus their Z-value (x-axis); the red dots show the estimated FDRs (y-axis) versus their Z-value (x-axis); and the blue stars show the BH adjusted p-values (y-axis) versus their Z-value (x-axis). This is the comparable plot to Figure \ref{fig1}, where the x-axis has been changed from p-value ranking to z-scale. The usefulness of this plot is that is shows what the desired FDR quantity is for a given Z-value. This provides context for our FDRs and adjusted p-values. 

\FloatBarrier

Here we see that Z-values greater than 2.85 and less than -2.5 have adjusted p-values less than 0.06 (blue threshold line, horizontal). This means in order to control the group-wise FDR, one would identify features with these Z-values as ``interesting". Notice that the Z-value above 4 has a FDR less than 0.06. Also the Z-value of 2.9 has a FDR less than 0.06. In practice, we find that the display in Figure \ref{fig1} is more intuitive for non-statisticians, but that Figure \ref{fig4} provides some essential insight into the stability and smoothness of the estimation procedure. 

\subsection{Lower Bound on the FDR}

The previous section introduced an empirical Bayes estimator for the FDR, which has become one of the most popular estimates. However, there are many different approaches for estimating the FDR. We have found it helpful in practice to be able to benchmark the magnitude of the FDR under known conditions in order to provide a contrast for estimators that rely heavily on distributional assumptions. This lower bound can help to contextualize findings and illuminate differences masked by empirical assumptions.

Our preferred benchmark is a well-known lower bound on the posterior probability of the null (hypothesis) under a gaussian model. This lower bound depends only on the data for the feature or test of interest and it does not borrow strength across features (for better or worse). Hence, it can also be used when only a single test is performed, i.e. when only a single p-value is available. In our experience, the gaussian assumption tends to have minimal influence because sampling distributions tend to be symmetric. 

The lower bound arises as follows. Let the joint density of data from a single feature be $g(X_1,...,X_n | \theta)$ where $\theta$ is a parameter of interest. The likelihood function is $L_n(\theta) \propto g(x_1,...,x_n | \theta)$ and denote the maximum likelihood estimator as $\hat{\theta}_n$. Recall that the null hypothesis is $H_0:\theta=\theta_0$. Let $\pi_0=P(null)$ be the prior probability of the null and let $z$ be the observed test statistic of the null hypothesis. Then, a lower bound on the posterior probability, $ P(null | x_1,..., x_n)$, which is effectively the FDR, is given by 

\begin{equation} \label{eq7}
\begin{split}
P(null | x_1,...x_n) \geq \left(1 + \frac{L(\hat{\theta}_n)}{L(\theta_0)} \frac{\pi_1}{\pi_0}\right)^{-1} \approx \left(1+\exp{(z^2/2)}\frac{\pi_1}{\pi_0}\right)^{-1}
\end{split}
\end{equation}

The first inequality holds because $\int g(x_1,...x_n|\theta_1)h(\theta_1) d \theta_1 \ leq g(x_1,...x_n | \hat{\theta}_n) $ for all $\theta_1\sim h(\theta_1)$. Note that $\int h(\theta_1) d \theta_1 =\pi_1$ by definition. The second approximation comes from the general asymptotic behavior of a classical likelihood ratio test, where $-2\log{ \frac{L(\theta_0)} {L(\hat{\theta}_n)}} \sim \chi_1^2 = {[N(0,1)]}^2$ for one-dimensional parameters. This lower bound is similar to that derived and explored by Berger (1985). Our function uses default odds, $\pi_1/ \pi_0 =1$, reasonable in many circumstances, which easily can be changed. As the z-statistic approaches zero, the lower bound approaches 1/2, as would be expected.

For illustration, consider feature 4 in Table \ref{tab1}. Feature 4 has an observed p-value of $0.051$, but has a univariate gaussian lower bound on the FDR of 0.13 $=(1+\exp(1.951^2/2))^(-1)$. In this case the BH estimated FDR is 0.064, substantially below the lower bound. This discrepancy in estimates is due to differing underlying assumptions. In contrast, feature 2 has a p-value of 0.049 and FDR of 0.122, very close to its lower bound. Although feature 4 has nearly the same p-value as feature 2, its BH FDR is nearly half that of feature 2. The univariate gaussian lower bound is helpful for identifying when FDR estimates may be optimistic, as in the case above. Similarly, we see that the adjusted p-values can be much less than the lower bound, which is another reason why they should not be mistaken for FDR estimates.

\section{Null Proportion $(\pi_0)$ Estimation}

The proportion of truly null features $(\pi_0)$, also known as the mixing proportion, is an important component of the FDR estimate that can be a strong driver of the estimate. While generally not identifiable, reasonable estimates of $\pi_0$ can be obtained under certain conditions. Many of the popular FDR estimation routines take a conservative approach by setting $\pi_0=1$, which results in a larger, i.e. conservative, FDR estimates.

The default in \texttt{p.fdr} is to assume that $\pi_0=1$. However, users are able to set the null proportion to a particular value or specify an estimation routine to estimate $\pi_0$ from the data. Many methods have been proposed for estimating the mixing proportion $\pi_0$ in a two-component mixture. \texttt{p.fdr} includes several of these methods such as Storey, Meinshausen, Jiang, Nettleton, and Pounds (\cite{storey:2003, mein:2006, jiang:2008, nett:2006, pounds:2003}). In next section, we propose a new approach that we call ``Last Histogram Height". This new approach is simple, appears to have excellent performance over a wide range of scenarios, and less computationally intensive that Storey's approach, which is quite popular. An evaluation and comparison to existing approaches is described in the subsequent subsection.

\subsection{Last Histogram Height}

Under the null, a test statistic for a feature, say a Z-value, is standard normal. As such, the corresponding p-value has a uniform distribution over the unit interval. Therefore, if all the features were null, we would expect an empirical histogram of the observed p-values to be approximately flat. Moreover, we see that the distribution of non-null p-values tends to be shifted toward zero.

The ``Last Histogram Height" method uses the bin height of p-values near 1 to estimate the true proportion of null features. We rely on the assumption that larger p-values are more likely to be come from null features. Let bin heights be $H_1, H_2, ... ,H_B$, where $B$ is the total number of bins. When $B=m$ ($m$ is the number of features) and all features are null, we would expect $H_i \approx 1$ for all $i = 1, ... , B$. The caveat is estimating bin height is sensitive to the choice of bin width. However, we have found that Scott's normal reference rule tends to work very well for this method \citep{scott:1979}.

\begin{figure}[H]
\centering
  \includegraphics[width=0.7\linewidth]{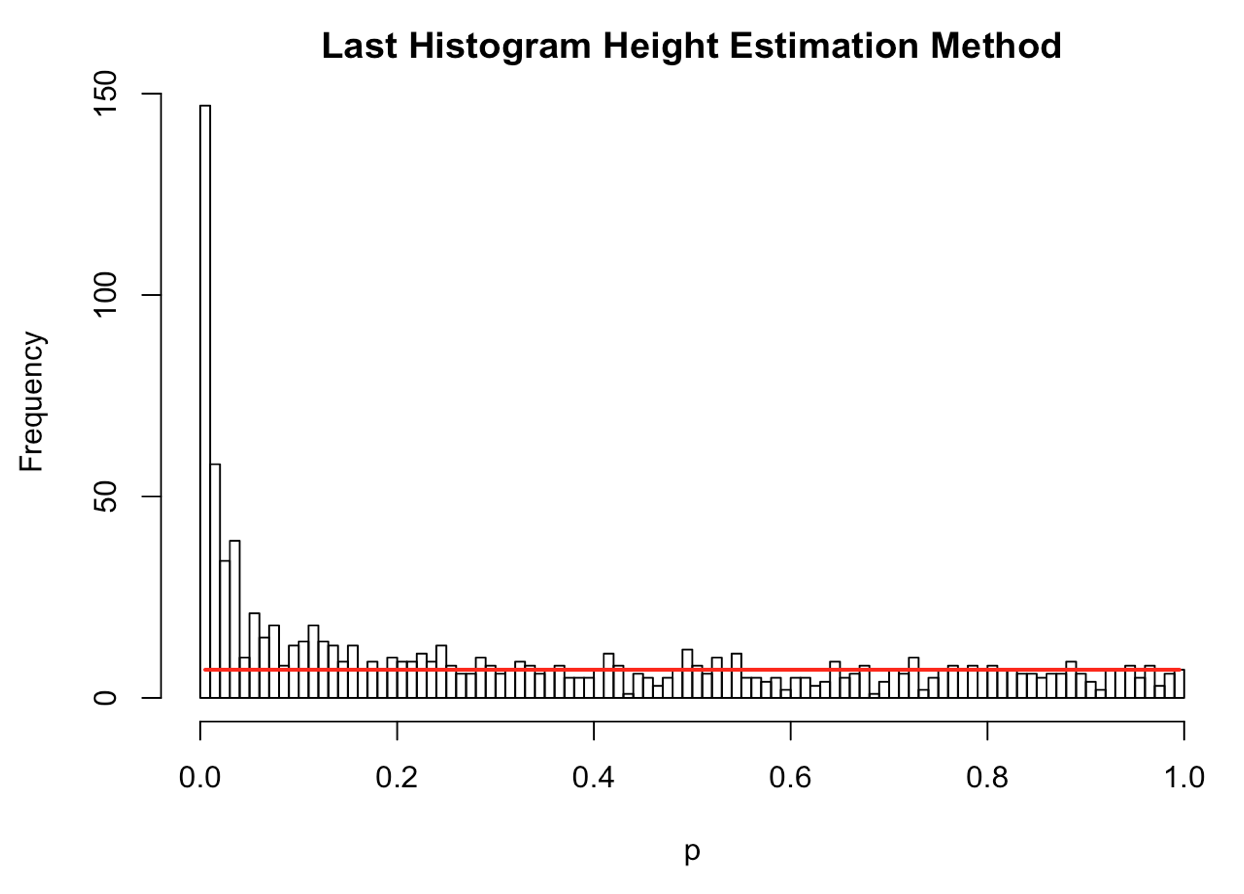}
  \caption{Simulated Example}
  \label{fig5}
\end{figure}

When $\pi_0<1$, the empirical distribution of the p-values (as shown by the histogram) will not be uniform over the unit interval. Departure from the uniform becomes easier to detect as $\pi_0$ moves further from 1 because the histogram shape quickly deviates from a uniform appearance. An example is presented in Figure \ref{fig5}, which shows a histogram of raw p-values from our simulated example of $m=1000$ features. The red horizontal line is drawn at the height of the last bin, $H_{B}$. In this approach $H_{B}$ is our ``null height'' and $H_{B} \cdot B$ is an estimate of the total number of null features. We then divide that by the number of total features ($m$) to estimate the null proportion (see Equation \eqref{eq8}):

\begin{equation} \label{eq8}
\begin{split}
\hat{\pi}_0=\frac{H_{B}B}{m}
\end{split}
\end{equation}

\FloatBarrier

The approach works because we would expect $\pi_0*m/B$ null p-values to be in each bin. This simple method performed well over many different simulation settings, as we described in the next section. It is also relatively free of constraining assumptions on the alternative distribution. We note that this approach can also be viewed as a form of central matching, as discussed by Efron \citep{efron:2013}, with center mass $\frac{1}{B}$ and a very small bin width. The ``Last Histogram Height" algorithm is as follows:

\begin{algorithm}[H]
\SetAlgoLined
\KwResult{Null proportion estimate}
\begin{enumerate}
\item Plot a histogram of the raw p-values, $p_1, p_2, ...p_m$, with $B$ number of bins, where $B<m$
\begin{itemize}
\item The most consistent bin method is \texttt{scott}, according to our simulations 
\end{itemize}
\item Store the histogram bin heights $H_{b}$ for each bin $b=1,2,...,B$
\item Call the height of last bin $H_{B}$ the ``null height'' 
\item Set the estimate of $\pi_0$ to be $$\hat{\pi}_0=\frac{H_{B}B}{m}$$
\end{enumerate}
 \caption{Last Histogram Height Method}
\end{algorithm}

\subsection{Storey}

Storey et al. (2003) propose an iterative procedure for estimating $\pi_0$. This procedure is popular and tends to have good performance characteristics over a wide range of scenarios. Storey's method relies on the fact that null p-values are uniformly distributed. As such, the bin height of p-values greater than 1/2 should give a conservative estimate of the null proportion. But there is nothing magical about 1/2, so Storey uses a tuning parameter. Let $\lambda$ identify ``large" p-values, e.g., $\#{p_i > \lambda}$ where $i=1,...,m$, such that the estimate of the null proportion $\hat{\pi}_0(\lambda)$, can be tuned by $\lambda$ to yield a desirable bias-variance tradeoff. Storey smoothes $\hat{\pi}_0(\lambda)$ before tuning, which provides some numerical stability. Note that for the ``Last Histogram Height" approach, the bin height closest to one is used to estimate the null proportion, which is conceptually similar to using $\lim_{\lambda \rightarrow 1} \hat{\pi}_0(\lambda)$ as Storey does. Storey's algorithm for estimating $\pi_0$ is as follows:

\newpage

\begin{algorithm}[H]
\SetAlgoLined
\KwResult{Null proportion estimate}
\begin{enumerate}
\item Let $p_{(1)},p_{(2)},...p_{(m)}$ be the ordered p-values. This also denotes the ordering of the features in terms of their evidence against the null hypothesis.
\item For a range of $\lambda$, say $\lambda=0, 0.05, 0.10, ... , 0.95$, and $i=1,..., m$, calculate $$\hat{\pi}_0(\lambda)=\frac{\# \{p_i>\lambda\}}{m(1-\lambda)}$$
\item Let $\hat{f}$ be the natural cubic spline with 3 df of $\hat{\pi}_0(\lambda)$ on $\lambda$
\item Set the estimate of $\pi_0$ to be when $\lambda =1$: $$\hat{\pi}_0=\hat{f}(1)$$
\end{enumerate}
 \caption{Storey's Method}
\end{algorithm}

\subsection{Comparison}

Below in Figure 6 are 3 plots showing the range of behavior of the 6 methods for estimating the null proportion that are included in our R package. These plots show the arrogate behavior of each method for estimating $\pi_0$ over 1000 simulations where the methods are used on a set of 100 features. A standard normal distribution was used for null features and 3 different alternative distributions were examined for alternative features (3 different plots). The x-axis represents the true $\pi_0$ used to generate data and ranges from 0 to 1.  The y-axis represents the average estimate $\pi_0$ (over the 1,000) simulations) for each of the 6 methods.  

"Last Histogram Height" and Storey's method preformed the best across these scenarios (and others not shown here). They routinely produce the closest estimates of the true null proportion. Although we only display 3 different mixture distributions for a set of 100 features here, we tested 12 different mixture distributions over 3 different features set sizes to confirm our results. We also tested the mean squared error and the results are well represented by the three examples given here. Our recommendation is to use the default of setting $\pi_0=1$ when the majority of features are expected to be null or nearly null. But in cases where the null proportion is likely to be different from one (say less than 0.95 or 0.9), then the ``Last Histogram Height" algorithm tends to perform the best.

\begin{centering}

\Large $\pi_0$ Estimates

\begin{multicols}{2}

\begin{figure}[H]
\centering
 \includegraphics[width=0.58\linewidth]{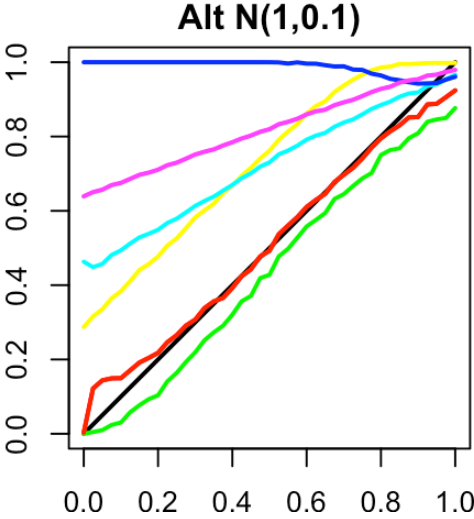}
\end{figure}

\begin{figure}[H]
\centering
 \includegraphics[width=0.58\linewidth]{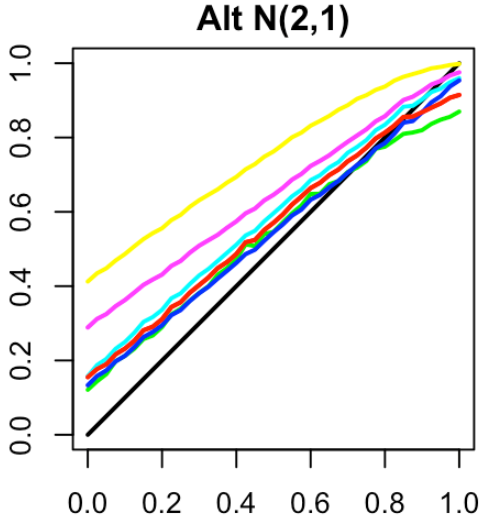}
\end{figure}

\end{multicols}

\begin{multicols}{2}

\begin{figure}[H]
\centering
 \includegraphics[width=0.58\linewidth]{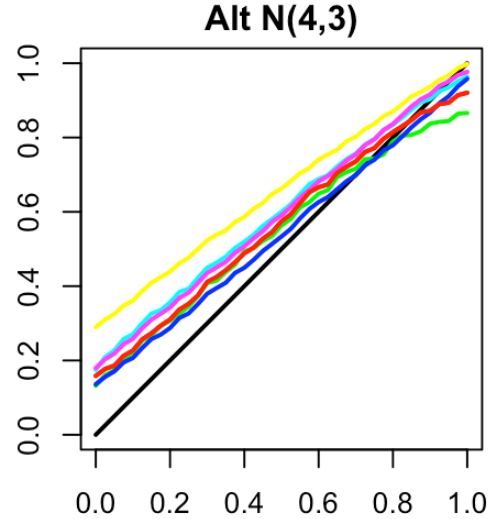}
\end{figure}

\begin{figure}[H]
\centering 
  \includegraphics[width=0.48\linewidth]{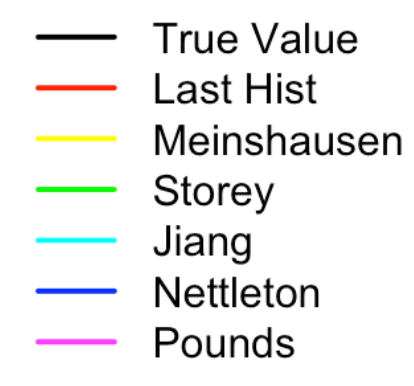}
\end{figure}

\end{multicols}
\label{fig6}

\normalsize
\textbf{Figure 6:} Comparison of Null Proportion Estimation methods

\end{centering}

\section{\texttt{FDRestimation} Package}

\texttt{FDRestimation} is a user-friendly R package that directly computes and displays false discovery rates from p-values or z-scores under a variety of assumptions. The following sections will explain the primary functions in this package and illustrate how to implement them. 

\subsection{\texttt{p.fdr} Function}

This \texttt{p.fdr} function is used to compute FDRs and multiple-comparison adjusted p-values from a vector of raw p-values. The stats package function \texttt{stats::p.adjust}  is similar in that it will produce multiple-comparison adjusted p-values. However, \texttt{stats::p.adjust} returns the BH adjusted p-value labeled as the FDR estimate. Strictly speaking this is inaccurate, because the BH FDR estimate should not have the forced monotonicity that its adjusted p-values must have. In addition, when estimating the FDR, our \texttt{FDRestimation::p.fdr}  function allows adjustments of key assumptions that are not adjustable in the \texttt{stats::p.adjust}  implementation (they are set to the simplest, most popular options).

This \texttt{FDRestimation::p.fdr}  function allows for the following adjustment methods: Benjamini-Hochberg, Benjamini-Yeukateili (with both positive and negative correlation), Bonferroni, Holm, Hochberg, and Sidak (\cite{bh:1995, by:2001, bon:1936, holm:1979, hoch:1988, sidak:1967}). It also allows the user to specify the threshold for important findings, the assumed $pi_0$ value, the desired $pi_0$ estimation method, whether to sort the results, and whether to remove NAs in the imputed raw p-value vector count (\texttt{stats::p.adjust}   actually counts NAs as viable features in its Bonferroni adjustment). Table \ref{tab2} shows all of the inputs for this function and their descriptions. 

The underlying methods for estimating the null proportion can be set by using the ``estim.method" and ``set.pi0" arguments. The default value of ``set.pi0" is 1, meaning it assumes that all features are null features. Accordingly, this approach will yield conservative estimates of the FDR. Alternatively, and less conservatively, one can attempt to estimate the null proportion from the data. To do this, we recommend using ``Last Histogram Height", as it was the simplest routine and one of the most accurate in our simulations. 

Here we see an example of how to use this \texttt{FDRestimation::p.fdr} function in R. We simulate 100 features with a true null proportion of 80$\%$. 

\begin{example}
set.seed(88888)

sim.data.p= c(runif(80),runif(20, min=0, max=0.01))

# Full set
p.fdr(p=sim.data.p, threshold=0.05, adjust.method="BH")

# First 5 p-values for Figure 7
p.fdr(p=sim.data.p[1:5], threshold=0.05, adjust.method="BH")

\end{example}

\begin{figure}[ht!]
\centering
  \includegraphics[width=0.7\linewidth]{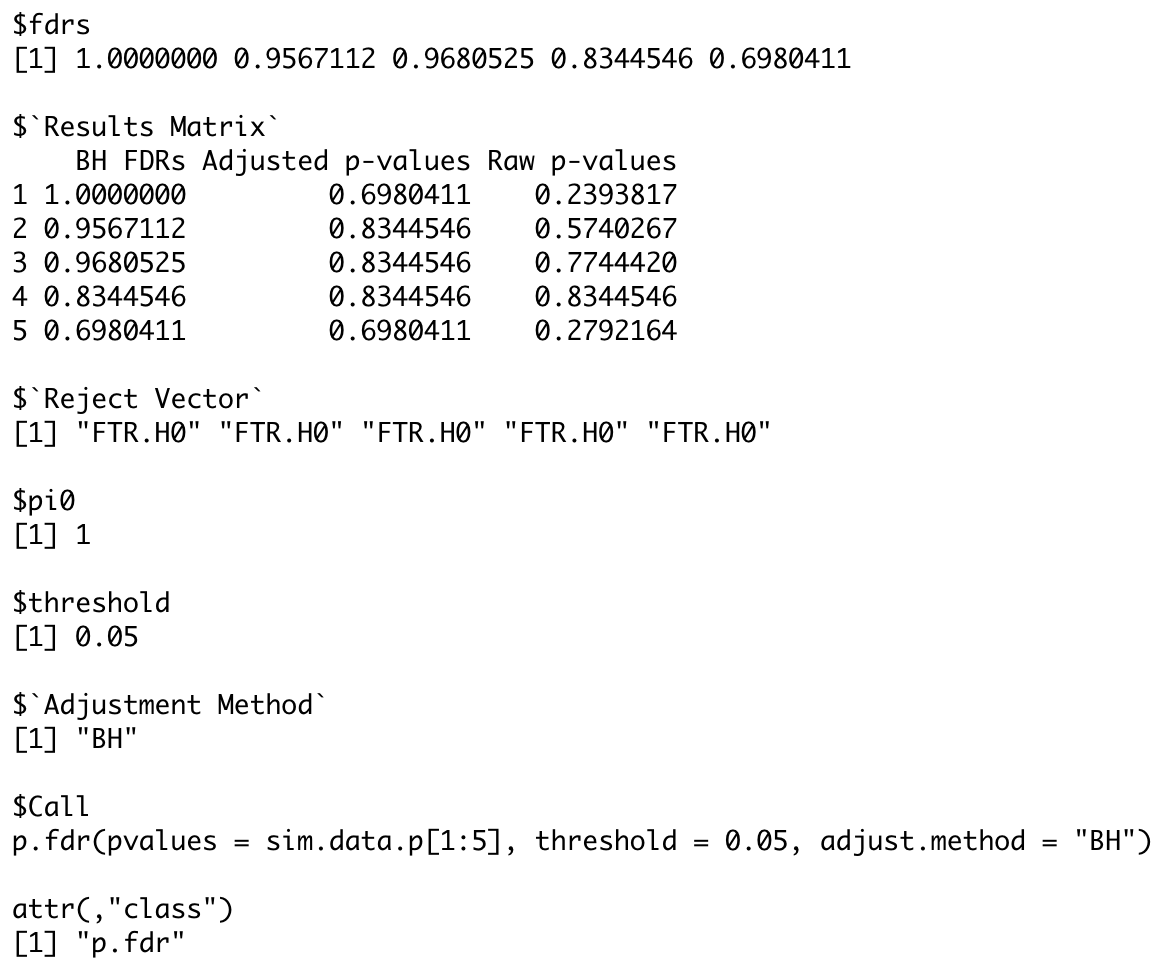}
  \caption{ }
  \label{fig7}
\end{figure}

The function will return a list object of the \texttt{p.fdr} class. In Figure \ref{fig7} we see this list object with the following components. 

\begin{itemize}

\item \textbf{fdrs}	A numeric vector of method adjusted FDRs.

\item \textbf{Results Matrix}	A numeric matrix of method adjusted FDRs, method adjusted p-values, and raw p-values.

\item \textbf{Reject Vector}	A vector containing Reject.H0 and/or FTR.H0 based off of the threshold value and hypothesis test on the adjusted p-values.

\item \textbf{pi0}	A numeric value for the pi0 value used in the computations.

\item \textbf{threshold}	A numeric value for the threshold value used in the hypothesis tests.

\item \textbf{Adjustment Method}	The string with the method name used in computation(needed for the plot.fdr function).
\end{itemize}

\FloatBarrier

\begin{table}[H]
\begin{tabular}{ll}
\hline
\multicolumn{1}{|l|}{\bf{Arguments}}     & \multicolumn{1}{l|}{\bf{Description}}\\ \hline

\multicolumn{1}{|l|}{pvalues}       & \multicolumn{1}{l|}{A numeric vector of raw p-values.}\\ \hline

\multicolumn{1}{|l|}{zvalues}       & \multicolumn{1}{l|}{A numeric vector of Z-values to be used in pi0 estimation }\\ 

\multicolumn{1}{|l|}{}       & \multicolumn{1}{l|}{or a string with options ``two.sided", ``greater" or ``less". Defaults to ``two.sided".}\\ \hline

\multicolumn{1}{|l|}{threshold}     & \multicolumn{1}{l|}{A numeric value in the interval $[0,1]$ used in a multiple comparisons}\\

\multicolumn{1}{|l|}{}       & \multicolumn{1}{l|}{hypothesis tests to determine significance from the null. Defaults to 0.05.} \\ \hline
 
\multicolumn{1}{|l|}{adjust.method} & \multicolumn{1}{l|}{A string used to identify the adjustment method. Defaults to \textit{BH}.} \\

\multicolumn{1}{|l|}{}       & \multicolumn{1}{l|}{ Options are \textit{BH}, \textit{BY}, \textit{Bon}, \textit{Holm}, \textit{Hoch}, and \textit{Sidak}.} \\ \hline

\multicolumn{1}{|l|}{BY.corr}       & \multicolumn{1}{l|}{A string of either ``positive" or ``negative" to determine which correlation}\\

\multicolumn{1}{|l|}{}       & \multicolumn{1}{l|}{ is used in the BY method. Defaults to \textit{positive}.}\\ \hline

\multicolumn{1}{|l|}{just.fdr}       & \multicolumn{1}{l|}{A Boolean TRUE or FALSE value which output only the FDR vector instead of }\\

\multicolumn{1}{|l|}{}       & \multicolumn{1}{l|}{ the list output. Defaults to FALSE.}\\ \hline

\multicolumn{1}{|l|}{default.odds}       & \multicolumn{1}{l|}{A numeric value determining the ratio of pi1/pi0 used in the computation of }\\

\multicolumn{1}{|l|}{}       & \multicolumn{1}{l|}{ single lower bound FDR. Defaults to 1.}\\ \hline

\multicolumn{1}{|l|}{estim.method}       & \multicolumn{1}{l|}{A string used to determine which method is used to estimate }\\

\multicolumn{1}{|l|}{}       & \multicolumn{1}{l|}{ the null proportion or pi0 value. Defaults to set.pi0.}\\ \hline

\multicolumn{1}{|l|}{set.pi0}       & \multicolumn{1}{l|}{A numeric value to specify a known or assumed pi0 value in the interval [0,1].  }\\

\multicolumn{1}{|l|}{}       & \multicolumn{1}{l|}{ Defaults to 1. Which means the assumption is that all inputted raw p-values come from }\\ 

\multicolumn{1}{|l|}{}       & \multicolumn{1}{l|}{ the null distribution.}\\ \hline

\multicolumn{1}{|l|}{hist.breaks}       & \multicolumn{1}{l|}{A numeric or string variable representing how many breaks are used in the pi0  }\\

\multicolumn{1}{|l|}{}       & \multicolumn{1}{l|}{ estimation histogram methods. Defaults to ``scott".}\\ \hline

\multicolumn{1}{|l|}{ties.method}       & \multicolumn{1}{l|}{A string a character string specifying how ties are treated. Options are "first", }\\

\multicolumn{1}{|l|}{}       & \multicolumn{1}{l|}{ "last", "average", "min", "max", or "random". Defaults to "random".}\\ \hline

\multicolumn{1}{|l|}{sort.results}       & \multicolumn{1}{l|}{A Boolean TRUE or FALSE value which sorts the output in either increasing or }\\

\multicolumn{1}{|l|}{}       & \multicolumn{1}{l|}{ non-increasing order dependent on the FDR vector. Defaults to FALSE.}\\ \hline

\multicolumn{1}{|l|}{na.rm}       & \multicolumn{1}{l|}{A Boolean TRUE or FALSE value indicating whether NA's should be removed from }\\

\multicolumn{1}{|l|}{}       & \multicolumn{1}{l|}{the inputted raw p-value vector before further computation. Defaults to TRUE.}\\ \hline
                                    &                                                                                                                                                                                                                                                                                 
\end{tabular}
\caption {\label{tab2} Inputs to the \texttt{p.fdr} function. }
\end{table}
\FloatBarrier

\subsection{\texttt{get.pi0} Function}

The \texttt{get.pi0} function is used to estimate the null proportion from the raw p-values. The user can choose one of 6 different methods included in our function: Last Histogram Height, Storey, Meinshausen, Jiang, Nettleton, and Pounds (\cite{storey:2003, mein:2006, jiang:2008, nett:2006, pounds:2003}). The user may also change the methods of determining the number of histogram breaks, which is an essential component for many of the methods implemented here. Table \ref{tab3} shows function arguments and their descriptions. 

Here we see an example of how to use this \texttt{get.pi0} function in R. We used the simulated data from above \texttt{sim.data.p} where the true null proportion was set to 80$\%$. In the first example, for the purposes of the estimation routine, $\pi_0$ was set to a single value with the \texttt{set.pi0=0.8} argument (1 is the default). Alternatively, we can use one of the 6 estimation methods in \texttt{get.pi0} instead of specifying $\pi_0$ a priori. Below is an example where we set the estimation method to \texttt{"last.hist"} (i.e., ``Last Histogram Height"). In that case, the get.pi routine returned an estimate of null proportion of 0.95. 

\begin{example}
set.seed(88888)

get.pi0(sim.data.p, estim.method="set.pi0", set.pi0=0.8)

[1] 0.8

get.pi0(sim.data.p, estim.method="last.hist")

[1] 0.85

\end{example}

\begin{table}[H]
\begin{tabular}{ll}
\hline
\multicolumn{1}{|l|}{\bf{Arguments}}     & \multicolumn{1}{l|}{\bf{Description}}\\ \hline

\multicolumn{1}{|l|}{pvalues}       & \multicolumn{1}{l|}{A numeric vector of raw p-values.}\\ \hline

\multicolumn{1}{|l|}{set.pi0}       & \multicolumn{1}{l|}{A numeric value to specify a known or assumed pi0 value in the interval [0,1].  }\\

\multicolumn{1}{|l|}{}       & \multicolumn{1}{l|}{ Defaults to 1. Which means the assumption is that all inputted raw p-values come from }\\ 

\multicolumn{1}{|l|}{}       & \multicolumn{1}{l|}{ the null distribution.}\\ \hline

\multicolumn{1}{|l|}{estim.method}       & \multicolumn{1}{l|}{A string used to determine which method is used to estimate }\\

\multicolumn{1}{|l|}{}       & \multicolumn{1}{l|}{ the null proportion or pi0 value. Defaults to set.pi0.}\\ \hline

\multicolumn{1}{|l|}{zvalues}       & \multicolumn{1}{l|}{A numeric vector of Z-values to be used in pi0 estimation }\\ 

\multicolumn{1}{|l|}{}       & \multicolumn{1}{l|}{or a string with options ``two.sided", ``greater" or ``less". Defaults to ``two.sided".}\\ \hline

\multicolumn{1}{|l|}{threshold}     & \multicolumn{1}{l|}{A numeric value in the interval $[0,1]$ used in a multiple comparisons}\\

\multicolumn{1}{|l|}{}       & \multicolumn{1}{l|}{hypothesis tests to determine significance from the null. Defaults to 0.05.} \\ \hline

\multicolumn{1}{|l|}{default.odds}       & \multicolumn{1}{l|}{A numeric value determining the ratio of pi1/pi0 used in the computation of }\\

\multicolumn{1}{|l|}{}       & \multicolumn{1}{l|}{ single lower bound FDR. Defaults to 1.}\\ \hline

\multicolumn{1}{|l|}{hist.breaks}       & \multicolumn{1}{l|}{A numeric or string variable representing how many breaks are used in the pi0  }\\

\multicolumn{1}{|l|}{}       & \multicolumn{1}{l|}{ estimation histogram methods. Defaults to ``scott".}\\ \hline

\multicolumn{1}{|l|}{na.rm}       & \multicolumn{1}{l|}{A Boolean TRUE or FALSE value indicating whether NA's should be removed from }\\

\multicolumn{1}{|l|}{}       & \multicolumn{1}{l|}{the inputted raw p-value vector before further computation. Defaults to TRUE.}\\ \hline

                                    &                                                                                                                                                                                                                                                                                 
\end{tabular}
\caption {\label{tab3} Inputs for the \texttt{get.pi0} function.}
\end{table}
\FloatBarrier

\subsection{\texttt{plot.p.fdr} Function}

This \texttt{plot.p.fdr} function is used to plot the results of \texttt{p.fdr}. By default, the adjusted FDRs, adjusted p-values and raw p-values are plotted along with two threshold lines to help contextualize the points. Any combination of p-values and thresholds can be removed from the plot. The user can set the axis limits, the location of the legend, the title of the plot and the plotting symbols and colors. Table \ref{tab4} shows all the function arguments and their descriptions. 

\begin{figure}[ht!]
\centering
  \includegraphics[width=0.8\linewidth, height=0.6\linewidth]{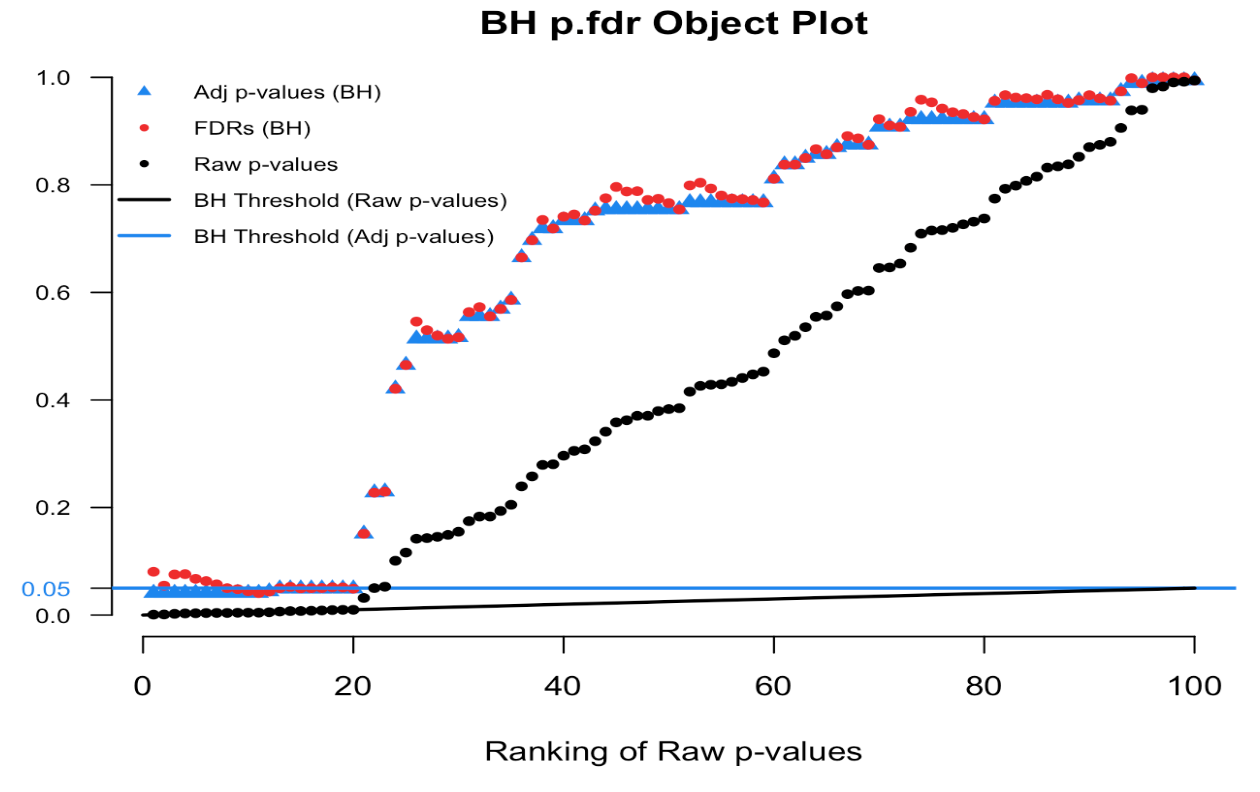}
  \caption{ }
  \label{fig8}
\end{figure}

\begin{figure}[ht!]
\centering
  \includegraphics[width=0.8\linewidth]{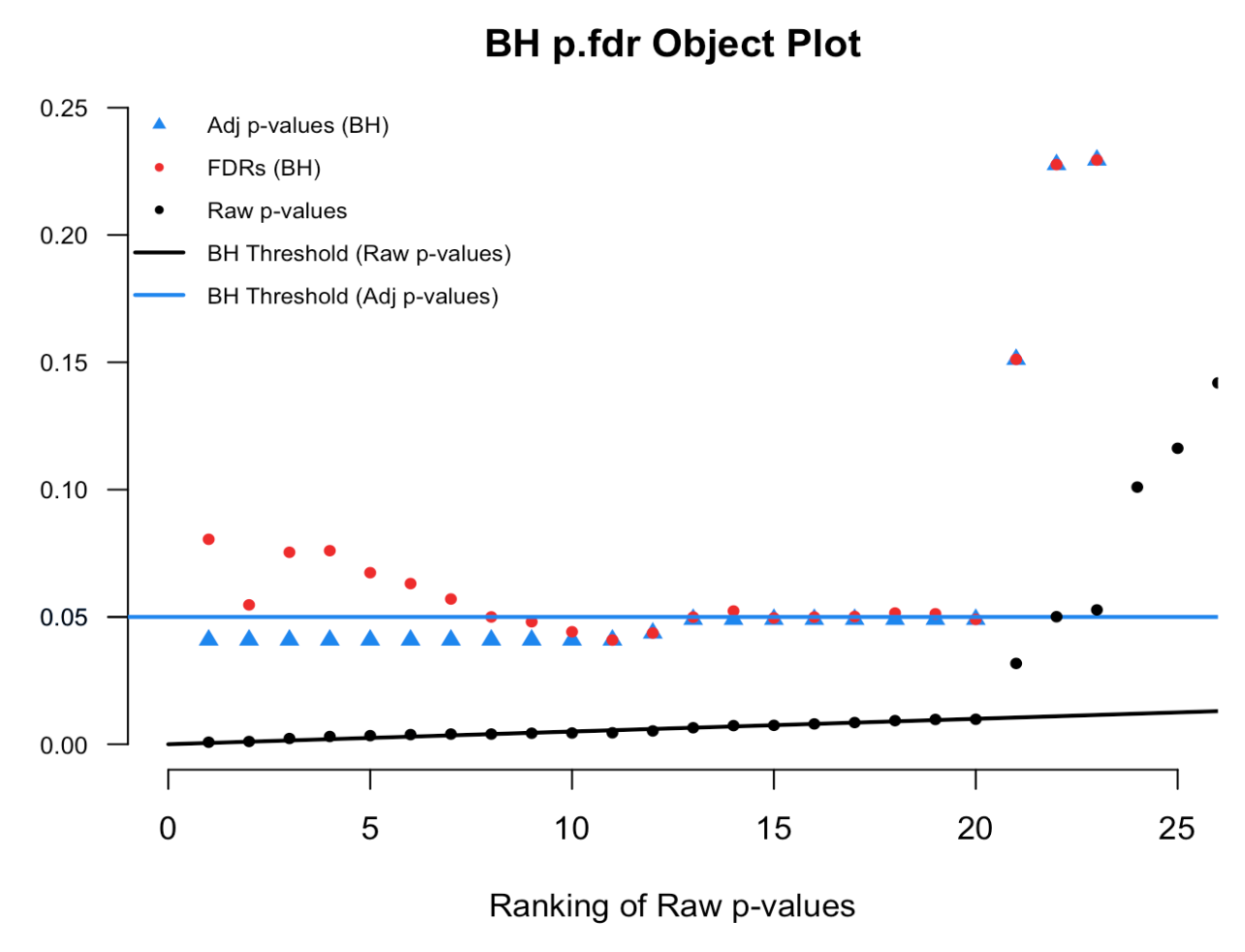}
  \caption{ }
  \label{fig9}
\end{figure}

Here we see an example of the \texttt{plot.p.fdr} function in R. We used our simulated data \texttt{sim.data.p}, where the a true null proportion was 80$\%$, for illustration. Figure \ref{fig8} show the default plot, and Figure \ref{fig9} zooms in on an interesting subset of findings. 

\begin{example}

# Figure 8
plot(p.fdr(p=sim.data.p))

# Figure 9
plot(p.fdr(p=sim.data.p), xlim=c(0,25), ylim=c(0,0.25))

\end{example}

\FloatBarrier

\FloatBarrier

\begin{table}[H]
\begin{tabular}{ll}
\hline
\multicolumn{1}{|l|}{\bf{Arguments}}     & \multicolumn{1}{l|}{\bf{Description}}\\ \hline

\multicolumn{1}{|l|}{p.fdr.object}       & \multicolumn{1}{l|}{A p.fdr object that contains the list of output.}\\ \hline

\multicolumn{1}{|l|}{raw.pvalues}       & \multicolumn{1}{l|}{A Boolean TRUE or FALSE value to indicate whether or not to plot the raw p-value }\\ 

\multicolumn{1}{|l|}{}       & \multicolumn{1}{l|}{points. Defaults to TRUE.}\\ \hline

\multicolumn{1}{|l|}{adj.pvalues}     & \multicolumn{1}{l|}{A Boolean TRUE or FALSE value to indicate whether or not to plot the adjusted }\\

\multicolumn{1}{|l|}{}       & \multicolumn{1}{l|}{p-value points. Defaults to TRUE.} \\ \hline
 
\multicolumn{1}{|l|}{sig.line} & \multicolumn{1}{l|}{A Boolean TRUE or FALSE value to indicate whether or not to plot the raw p-value} \\

\multicolumn{1}{|l|}{}       & \multicolumn{1}{l|}{ significance line. Defaults to TRUE.} \\ \hline

\multicolumn{1}{|l|}{adj.sig.line}       & \multicolumn{1}{l|}{A Boolean TRUE or FALSE value to indicate whether or not to plot the adjusted }\\

\multicolumn{1}{|l|}{}       & \multicolumn{1}{l|}{ significance threshold. Defaults to TRUE.}\\ \hline

\multicolumn{1}{|l|}{threshold}     & \multicolumn{1}{l|}{A numeric value in the interval $[0,1]$ used in a multiple comparisons}\\

\multicolumn{1}{|l|}{}       & \multicolumn{1}{l|}{hypothesis tests to determine significance from the null. Defaults to 0.05.} \\ \hline

\multicolumn{1}{|l|}{x.axis}       & \multicolumn{1}{l|}{A string variable to indicate what to plot on the x-axis. Can either be ``Rank" or  }\\

\multicolumn{1}{|l|}{}       & \multicolumn{1}{l|}{ ``Zvalues". Defaults to ``Rank".}\\ \hline

\multicolumn{1}{|l|}{xlim}       & \multicolumn{1}{l|}{A numeric interval for x-axis limits. }\\ \hline

\multicolumn{1}{|l|}{ylim}       & \multicolumn{1}{l|}{A numeric interval for y-axis limits. Defaults to c(0,1). }\\ \hline

\multicolumn{1}{|l|}{zvalues}       & \multicolumn{1}{l|}{A numeric vector of Z-values to be used in pi0 estimation }\\ 

\multicolumn{1}{|l|}{}       & \multicolumn{1}{l|}{or a string with options ``two.sided", ``greater" or ``less". Defaults to ``two.sided".}\\ \hline

\multicolumn{1}{|l|}{legend.where}       & \multicolumn{1}{l|}{A string ``bottomright", ``bottomleft", ``topleft", ``topright". Defaults to ``topleft" }\\

\multicolumn{1}{|l|}{}       & \multicolumn{1}{l|}{  is x.axis="Rank" and ``topright" if x.axis="Zvalues".}\\ \hline

\multicolumn{1}{|l|}{main}       & \multicolumn{1}{l|}{A string variable for the title of the plot. }\\ \hline

\multicolumn{1}{|l|}{pch.adj.p}       & \multicolumn{1}{l|}{A plotting 'character', or symbol to use for the adjusted p-value points. This can 
}\\

\multicolumn{1}{|l|}{}       & \multicolumn{1}{l|}{either be a single character or an integer code for one of a set of graphics symbols.}\\ 

\multicolumn{1}{|l|}{}       & \multicolumn{1}{l|}{Defaults to 17.}\\ \hline

\multicolumn{1}{|l|}{pch.raw.p}       & \multicolumn{1}{l|}{A plotting 'character', or symbol to use for the raw p-value points. This can either 
}\\

\multicolumn{1}{|l|}{}       & \multicolumn{1}{l|}{be a single character or an integer code for one of a set of graphics symbols. }\\ 

\multicolumn{1}{|l|}{}       & \multicolumn{1}{l|}{Defaults to 20.}\\ \hline

\multicolumn{1}{|l|}{pch.adj.fdr}       & \multicolumn{1}{l|}{A plotting 'character', or symbol to use for the adjusted FDR points. This can 
}\\

\multicolumn{1}{|l|}{}       & \multicolumn{1}{l|}{either be a single character or an integer code for one of a set of graphics symbols. }\\ 

\multicolumn{1}{|l|}{}       & \multicolumn{1}{l|}{Defaults to 20.}\\ \hline

\multicolumn{1}{|l|}{col}       & \multicolumn{1}{l|}{A vector of colors for the points and lines in the plot. If the input has 1 value all  
}\\

\multicolumn{1}{|l|}{}       & \multicolumn{1}{l|}{points and lines will be that same color. If the input has length of 3 then }\\ 

\multicolumn{1}{|l|}{}       & \multicolumn{1}{l|}{col.adj.fdr will be the first value, col.adj.p will be the second, and col.raw.p is the third. }\\ 

\multicolumn{1}{|l|}{}       & \multicolumn{1}{l|}{Defaults to c("dodgerblue","firebrick2", "black"). }\\ \hline
                                    &                                                                                                                                                                                                                                                                                 
\end{tabular}
\caption {\label{tab4} Inputs for the \texttt{plot.p.fdr} function.}
\end{table}
\FloatBarrier

\section{Adjustment Methods}

The computation of adjusted p-values and FDRs for each method follows a similar intuitive approach. First, estimates of the FDR for each feature are obtained using the preferred method, e.g. Benjamini-Hochberg or Benjamini-Yekutieli. Step-up or step-down adjustments are not applied at this stage. Next, adjusted p-values are obtained from the estimated FDRs by applying the step-up or step-down adjustment that is associated with the method. The step adjustment is necessary for error control but not for FDR estimation. For methods that do not have a step-up or step-down component, e.g. Bonferroni, the adjusted p-values and FDRs will be the same. The distinction between the estimated FDRs and the adjusted p-values is an important one that is routinely confused in practice.

Note that all estimates of adjusted p-values and FDRs are forced to be 1 or less. Also, when ranks are used in our package the \texttt{ties.method = "random"}. This means for example that if the 4 smallest p-values in a vector tie in value then they will be assigned ranks 1,2,3,4 randomly. The user can change the ties method in the input to the function. 

Below we illustrate this with the remaining 5 methods (BH is discussed above).

\subsection{Benjamini-Yekutieli}

Benjamini-Yekutieli (BY) is a step-up method for controlling the false discovery rate under arbitrary dependence \citep{by:2001}. For a pre-specified dependence structure, there exits an adjustment function called $c(m)$ that is used to modify the Benjamini-Hochberg estimate of the FDR. For example, in the case of flexible positive dependence, the function $ c(m)=\sum_{j=1}^m \frac{1}{j}$ is used. Then, the thresholding criteria is to find the largest index $i$ such that

\begin{equation} \label{eq9}
\begin{split}
p_{(i)} \leq \gamma \frac{i}{m} \frac{1}{c(m)}   
\end{split}
\end{equation}

which is a scaled version of the BH criterion given in Equation \eqref{eq1}. 

This can be written compactly $k=\max{\left[i: p_{(i)} \leq \gamma \cdot i / (m \cdot c(m))  \right]}$ or for non-ordered vectors of p-values $k=\max{\left[\text{rank} (p_i): p_{i} \leq \gamma \cdot \text{rank} (p_i) / (m \cdot c(m))  \right]}$. Then all features with $p_{(1)},...,p_{(k)}$ are deemed interesting at the FDR $\gamma$ threshold and considered ``findings".
Recall that Benjamini-Hochberg procedure uses the step function ($F(p_{(i)})=i/m$) as its implicit empirical estimate of the mixing distribution function (CDF) Check this notation. The Benjamini-Yekutieli procedure amounts to simply using a modified estimate for the CDF, namely ($F(p_{(i)})=i/(m \cdot c(m))  $). 

Mathematically, the adjusted p-values and estimated FDRs are

\begin{equation} \label{eq10}
\begin{split}
\tilde{p}^{BY}_{(i)} \coloneqq \min_{j \geq i}\left(p_{(j)} \frac{ m\cdot c(m)}{j}\right) \leq \gamma
\end{split}
\end{equation}

\begin{equation} \label{eq11}
\begin{split}
FDR^{BY}_i \coloneqq p_i \frac{ m \cdot c(m)}{\text{rank}(p_i)} \cdot \hat{\pi}_0
\end{split}
\end{equation}

Comparing this form to the general formula for the FDR in Equation \eqref{eq6}, we see that the BY correction amounts to changing the estimate of the mixture distribution $F(\mathcal{Z})$ from $[\text{rank}(p_i)/m]$ to $[\text{rank}(p_i)/ (m \cdot c(m))]$ to account for dependence. Note that we have avoided using the ordered notation for False discovery rate estimates, say $ FDR_{(i)}$, because although those estimates are dependent on ordered p-values the FDR estimates themselves do not have to be monotonic.

\begin{figure}[ht!]
\centering
  \includegraphics[width=0.8\linewidth]{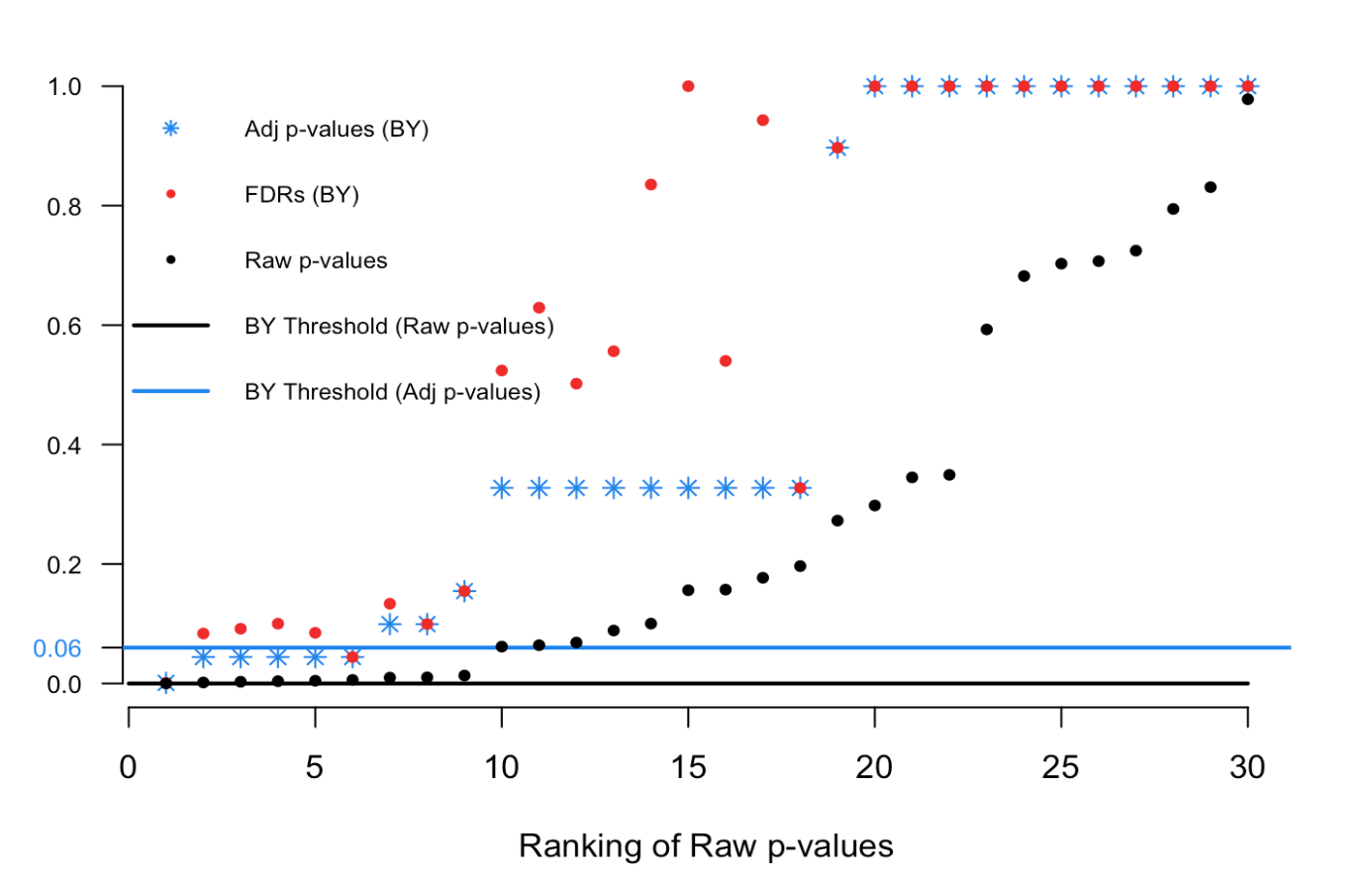}
  \caption{ }
  \label{fig10}
\end{figure}
\FloatBarrier

Here we see the BY FDRs, or red points, jump above and below the 0.06 threshold in ranks 1 to 6. Then in ranks 7 and greater the red dots remain above the threshold and quickly are adjusted to the value of 1. The positive dependence correction causes these BY FDRs to be closer to 1, or more conservative. 

\subsection{Bonferroni}

The Bonferroni correction controls the family wise error rate (FWER) \citep{bon:1936}. We include it in our function because of its popularity in multiple adjustments even though it is not directly related to FDR. For this method we would reject the null hypothesis for each $p_i \leq \frac{\gamma}{m}$ in order to control the FWER at $\leq \gamma$ level. In our functions the adjusted p-values and adjusted FDRs will always be identical for this method. 

\begin{equation} \label{eq12}
\begin{split}
\tilde{p}^{Bon}_{i} \coloneqq p_im  \leq \gamma
\end{split}
\end{equation}

\begin{equation} \label{eq13}
\begin{split}
FDR^{Bon}_i \coloneqq p_im \cdot \hat{\pi}_0
\end{split}
\end{equation}

From this form we see that the Bonferroni correction amounts to changing the estimate of the mixture distribution $F(\mathcal{Z})$ to $[1/ m]$. 

\subsection{Sidak}

The Sidak or Dunn-Sidak correction controls the family wise error rate (FWER) \citep{sidak:1967}. This correction method is exact for tests that are independent, it is conservative for tests that are positively dependent, and it is liberal for tests that are negatively dependent. For this method is slightly less strict then the traditional Bonferroni method. For each $p_i \leq \gamma_{Sid} = 1-(1-\gamma)^{\frac{1}{m}}$ reject the null hypothesis in order to control the FWER at $\leq \gamma$ level. In our functions the adjusted p-values and adjusted FDRs will always be identical for this method. 

\begin{equation} \label{eq14}
\begin{split}
\tilde{p}^{Sid}_{i} \coloneqq 1-(1-p_i)^{m}  \leq \gamma
\end{split}
\end{equation}

\begin{equation} \label{eq15}
\begin{split}
FDR^{Sid}_i \coloneqq 1-(1-p_i)^{m} \cdot \hat{\pi}_0
\end{split}
\end{equation}

From this form we see that the Sidak correction amounts to changing the estimate of the mixture distribution $F(\mathcal{Z})$ to $[p_i/(1-(1-p_i)^{m})]$ assuming $F_0(Z)=p_i$. 

\subsection{Holm}

The Holm method, also known as the Holm-Bonferroni method, controls the FWER and is less conservative and therefore uniformly more powerful than the Bonferroni correction \citep{holm:1979}. For this method we use the step-down procedure which would reject the null for those rankings $1,...,(k-1)$ such that $k$ is the smallest ranking where: 

\begin{equation} \label{eq16}
\begin{split}
p_{(k)} \leq \frac{\gamma}{m+1-k}
\end{split}
\end{equation}

From the above equation we see that it relies on the ranking or $j$ that means our function's outputted adjusted p-value and FDR can be different. 

\begin{equation} \label{eq17}
\begin{split}
\tilde{p}^{Holm}_{(i)} \coloneqq \max_{j \leq i}\left(p_{(j)}(m+1-j)\right) \leq \gamma
\end{split}
\end{equation}

\begin{equation} \label{eq18}
\begin{split}
FDR^{Holm}_i \coloneqq p_i(m+1-\text{rank}(p_i)) \cdot \hat{\pi}_0
\end{split}
\end{equation}

From this form we see that the Holm correction amounts to changing the estimate of the mixture distribution $F(\mathcal{Z})$ to $[1/ (m+1-\text{rank}(p_i))]$.

\subsection{Hochberg}

The Hochberg method uses the same equation as the Holm method, Equation \eqref{eq16} \citep{hoch:1988}. However for this method we use the step-up procedure. This means we would reject the null for those rankings $1,...,j$ such that $j$ is the largest ranking where: 

\begin{equation} \label{eq19}
\begin{split}
p_{(j)} \leq \frac{\gamma}{m+1-j}
\end{split}
\end{equation}

This change from the step-down to the step-up procedure results in the Hochberg correction being more powerful than the Holm method. 

\begin{equation} \label{eq20}
\begin{split}
\tilde{p}^{Hoch}_{(i)} \coloneqq \min_{j \geq i}\left(p_{(j)}(m+1-j)\right) \leq \gamma
\end{split}
\end{equation}

\begin{equation} \label{eq21}
\begin{split}
FDR^{Hoch}_i \coloneqq p_i(m+1-\text{rank}(p_i)) \cdot \hat{\pi}_0 
\end{split}
\end{equation}

From this form we see that the Hochberg correction is the same as the Holm and amounts to changing the estimate of the mixture distribution $F(\mathcal{Z})$ to $[1/ (m+1-\text{rank}(p_i))]$.

%
%
%
%
%

\section{Comments}

We encourage the use of FDR methods and desire to illuminate the importance of contextualizing important findings. Our package provides useful and easy tool for those want to compute the false discovery rate, analogous to the role that \texttt{stats::p.adjust} plays for multiple comparison adjustments in everyday practice. Importantly, we hope it is now clear that p-value adjustments are not interchangeable with FDRs. In addition, \texttt{FDRestimation} package clearly delineates between methods for FDR control and methods for FDR estimation, while still allowing the user to choose from many different inputs and assumptions for their data. The more flexibility the user has at their disposal with these methods, better interpretations and applications will result.

\bibliography{Rough_Draft_wrapper}

\address{Megan Hollister Murray\\
  Vanderbilt University\\
  Department of Biostatistics\\
  Nashville, TN\\
  \email{megan.c.hollister@vanderbilt.edu}}

\address{Jeffrey D. Blume\\
  Vanderbilt University\\
  Department of Biostatistics\\
  Nashville, TN\\
  \email{j.blume@vanderbilt.edu}}

%
%
 
%
%
%
%
%
%
%
%
%
%
%
%
%
%
%

\end{article}

\end{document}